# Massart iron oxide nanoparticles in mechanobiology

**Myriam Reffay[1], Gilles Tessier[2] and Jean-François Berret[1*]**

[1]*Université Paris Cité, CNRS, Matière et systèmes complexes, 75013 Paris, France*
[2] *Sorbonne Université, INSERM, CNRS, Institut de la Vision, 17 rue Moreau, F-75012, Paris, France*

**Abstract**
Magnetic nanoparticles (MNPs) derived from the Massart coprecipitation method have played a pioneering role in bridging materials science and biology. Their magnetic moment and nanoscale dimensions have long enabled applications in magnetic resonance imaging, targeted drug delivery, separation technologies, and hyperthermia. Beyond these well-established uses, a growing research direction has emerged at the interface of physics and biology: the application of MNPs in mechanobiology, the study of how mechanical forces regulate cellular and tissue functions. This review examines how Massart MNPs can be used to generate, transmit, and measure forces within living systems. Particular attention is given to interfacial control through advanced surface chemistries, which ensure colloidal stability, minimize toxicity, and preserve the nanoscale integrity of the particles. We also show that the robustness and scalability of Massart synthesis make it ideally suited to the production of biocompatible nanomaterials in quantities required for comprehensive biological studies. Two complementary approaches are discussed. The first exploits MNP assemblies that arise spontaneously within cells, in the form of endosomes. These compartments enable the application of controlled magnetic forces to study both the formation of reconstructed tissues and organoids, and their viscoelastic response. The second focuses on micrometric magnetic wires fabricated from MNPs, used in active microrheology to probe the cytoplasm of living cells over a wide frequency range. Together, these approaches illustrate how MNP assemblies can accurately quantify cellular and tissue mechanics, providing new opportunities for mechanobiological studies and enabling the development of novel readouts of cellular mechanics with potential diagnostic applications.

**Keywords:** Iron oxide nanoparticles | Massart synthesis | Mechanobiology| Tissue engineering | Cellular aggregates | Microrheology | Viscosity mapping | Heterodyne holography | Cancer cells

# I – Introduction

Magnetic colloidal dispersions, historically known as ferrofluids, are liquids containing iron oxide nanocrystals, most often magnetite ($Fe_3O_4$) or maghemite ($\gamma$-$Fe_2O_3$), that has the ability to react to an applied magnetic field. In his seminal articles published in the early 1980s,[1,2] Massart introduced a simple and robust chemical process based on the coprecipitation of iron salts in aqueous solution, enabling the preparation of stable dispersions without the use of organic stabilizing agents. Because of its simplicity, reproducibility, and scalability, the Massart coprecipitation method rapidly became a benchmark in inorganic chemistry and was later exploited in other disciplines, including physics, medicine, nanotechnology, and more recently biology.[3-6] One of the key features of this synthesis is that it is performed in an aqueous medium and allows the preparation, in a single batch, of large quantities of well-dispersed nanoparticles typically 5–20 nm in size. Furthermore, the particles remain within the paramagnetic domain,[7,8] a unique characteristic whose advantages will be discussed in this review. This method has become so widely adopted that many publications still employing the coprecipitation of iron salts no longer cite Massart original work, underscoring how this synthesis has become a standard protocol for nanoparticle preparation. The present contribution to this special issue also aims to pay tribute to René Massart pioneering role in the development of modern soft chemistry.

In the vast literature on magnetic nanoparticles, numerous acronyms have been introduced. Common examples include IONP for iron oxide nanoparticle,[9-11] SPION for superparamagnetic iron oxide nanoparticle,[11,12] USPIO (or VSOP) for ultrasmall (or very small) superparamagnetic iron oxide,[5,13] and MION for monocrystalline iron oxide nanoparticle.[14,15] These designations do not refer to a specific synthesis method, and also cover particles obtained by the Massart method. More recently, the concise abbreviation MNPs, for magnetic nanoparticles, has become prevalent.[16,17] In this review, the term Massart magnetic nanoparticles will be used, abbreviated as MNPs. When referring to MNPs not produced by the Massart synthesis, this will be indicated.

Since their initial synthesis, MNPs have been employed across a broad spectrum of research areas and technological applications. Their use spans ferrohydrodynamics, energy conversion and heat transfer, magnetic data storage and recording, sensors, and the controlled assembly of nanostructures for the fabrication of smart, stimulus-responsive materials.[3,18-21] In life sciences, the earliest studies explored their potential as contrast agents for magnetic resonance imaging (MRI),[5] as tools for magnetic separation and purification of biomolecules,[22] and as components in biosensing, targeted drug delivery, and magnetic hyperthermia for cancer therapy.[16,17,23,24] A particularly successful development has been their incorporation into polymer beads for magnetic separation technologies,[22] or into elastomer matrices to produce hybrid materials that combine mechanical flexibility with magnetic responsiveness.[24,25]

Compared with the applications mentioned above, MNPs have been relatively little explored in mechanobiology. Mechanobiology is the field that investigates how the mechanical properties of cells and tissues regulate biological functions such as differentiation, migration, and proliferation.[26,27] Cellular components including cytoskeleton, organelles, and plasma membrane continuously sense and respond to mechanical stimuli from their environment. Understanding these processes requires integrating local mechanical measurements with the underlying biological mechanisms. These measurements are performed using advanced microrheology techniques[28-30] that enable the quantification of viscoelastic properties relevant to *in vivo* living systems.[31] These include the intracellular scale (below 10 μm), the single-cell scale (10–30 μm), and the tissue scale (above 30 μm).[31,32] In this review, we highlight two complementary approaches based on Massart particles: one applying magnetic forces at the tissue scale, and the other generating magnetic torques at the intracellular level.

The use of MNPs in mechanobiology faces, however, inherent limitation: the magnetic forces that can be exerted on single MNPs are extremely weak. This limitation arises from two main factors. First, the particles are only a few nanometers in size, and their magnetic moments are therefore very small, of the order of $10^{-19}$ A m². Secondly, in many experimental setups used in cellular biomechanics, applying magnetic field gradients that are both strong enough to exert significant forces on individual MNPs and still compatible with cellular mechanics is challenging. For instance, a single 10 nm MNP exposed to a typical laboratory field gradient experiences a force on the order of $10^{-18}$ newtons. In this case, the particle magnetic energy is lower than, or on the order of, the thermal energy $k_B T$.[7,21,33] By comparison, the characteristic mechanical forces involved in biological processes, such as cell–cell adhesion, cell–substrate traction, or tissue contractility, range from nano- to piconewtons.[26,34,35]

In this review, we show that this limitation can be overcome by exploiting collective (rather than individual) MNP behaviors. The particles can be assembled into large magnetic structures, typically a few micrometers in size and containing up to several million MNPs. These assemblies may take the form of intracellular vesicles, such as densely loaded endosomes, or synthetic aggregates within elastomeric matrices.[22,24,36-39] Throughout this review, we refer to these structures as *MNP assemblies*. In such systems, the magnetic force scales with the total magnetic volume, enabling the generation of forces in the piconewton range for $10^6$ MNPs, which is comparable to the forces naturally produced by cells.

This review explores the use of Massart iron oxide nanoparticles and their assemblies in mechanobiology. The first part examines the MNP physicochemical, and magnetic properties, highlighting the critical importance of their surface coating in the context of biological environments. The second part focuses on individual MNPs internalized by cells, where they reside within the cytoplasm in membrane-bound compartments, defining the first instance of MNP assemblies. Here, we investigate the conditions under which cells in suspension exposed to external magnetic fields self-organize into tissue-like structures, and examine how forces can be applied to measure the viscosity and elasticity of whole cell aggregates. The final part addresses micrometric MNP assemblies such as magnetic wires, which can be magnetically actuated to generate torque and serve as intracellular probes for measuring cytoplasmic viscoelasticity. The review concludes with recent photonic advances for mapping viscosity of model fluids and the application of cancer cell microrheology to assess their metastatic potential.

## II – Massart synthesis, magnetic and physicochemical properties

### II.1 - Physical chemistry of Massart iron oxide nanoparticles

Massart original synthesis was first described in the *Comptes rendus de l'Académie des sciences* (Paris) in 1980 and later published in English in the proceedings of the 2nd IEEE Conference on Magnetism held in Orlando that same year.[1,2] Inspired by Elmore work four decades earlier[40] Massart developed a method to obtain stable ferrofluids in aqueous media without the use of organic stabilizing agents. In Massart synthesis, colloidal dispersions of magnetite are prepared by the alkaline co-precipitation of ferrous and ferric chlorides, followed by peptization with either an acid ($HNO_3$) or a base ($N(CH_3)_4OH$, tetramethylammonium hydroxide). In acidic conditions (pH 2), the iron oxide surface bears a positive charge ($Fe-OH_2^+$) due to Brønsted acid–base equilibria, with nitrate ions as counterions.[41] In basic conditions (pH 12), the nanocrystal surfaces are negatively charged, with deprotonated $Fe-O^-$ groups. In both cases, subsequent oxidation in a ferric nitrate solution at 100 °C produces maghemite ($\gamma$-$Fe_2O_3$) nanocrystals with sizes in the range of 5–20 nm. For maghemite, the surface charge is determined by the medium acidity, following acid–base equilibria characterized by $pK_1 = 4.6$ and $pK_2 = 8.2$.[41]

$$Fe-OH_2^+ + OH^- \overset{K_1}{\leftrightarrows} Fe-OH + H_2O \overset{K_2}{\leftrightarrows} Fe-OH^- + H_3O^+ \qquad (1)$$

Size sorting by selective precipitation was subsequently introduced to narrow the particle size distribution.[42,43] Fig. 1a shows maghemite dispersions obtained by Massart synthesis at increasing MNP concentrations orange-yellow colors below 1 g $L^{-1}$ and shifting to a garnet-red color at high concentrations. The corresponding mass absorptivity in Fig.1b reveal a pronounced increase in absorbance below 600 nm and a characteristic shoulder at 380 nm (arrow).

Owing to its simplicity, the Massart synthesis method effectively meets the essential requirements for conducting experiments in mechanobiology, which is the central theme of this review. It allows the preparation of large amounts of nanoparticles in stable dispersions, typically 1–10 g dry matter in a single synthesis. The controlled size range (5–20 nm) ensures superparamagnetic behavior, enabling predictable magnetization and force application based on paramagnetism theory.[7,21] Fig. 1c and 1d show transmission electron microscopy (TEM) images of MNPs obtained by Massart synthesis combined with size sorting, with diameters of 6.2 and 11.0 nm, respectively. The relative frequency plots in the right-hand panels display the particle size distributions, with dispersity indices $s \simeq 0.20$, where $s$ is defined as the ratio of the standard deviation to the median value of the distribution.[44] The particles exhibit in addition high surface charge, cationic at acidic pH or anionic at alkaline pH, providing electrostatic stabilization of the sol.[45] Under physiological conditions, the particles are close to electrical neutrality and therefore tend to aggregate under the influence of van der Waals attractions. The structural charges present at the nanocrystal surface can be used promote non-covalent interactions with organic (macro)molecules such as ligands or polymers, enabling the formation of a stable coating.

## II.2 - Coating strategies

MNPs biological applications emerged in the 1990s and focused on cellular therapies, tissue formation and repair, drug delivery, MRI, and hyperthermia.[5,10,13] The case of MRI is particularly revealing, as it underscores the importance of controlling the interface between nanocrystals and complex biological environments. Massart particles can act as MRI contrast agents by generating local magnetic fields that modify the relaxation rates of nearby hydrogen nuclei, thereby altering the signal intensity in the body regions where they are present. For MRI contrast agents, various coating strategies were developed including macromolecules derived from dextran or PEGylated starch.[5] During that period, commonly used clinical contrast agents were dextran-coated formulations marketed under the names of Endorem® (Berlex Laboratories), Sinerem® (Guerbet), and Resovist® (Schering).[13] While these coatings were effective in stabilizing dispersions ex vivo, they were rapidly recognized and cleared by the immune system *in vivo*, particularly through hepatic uptake. As a result, these agents were ultimately restricted to liver imaging.[5] These outcomes were ascribed to non-specific interfacial bonding between glucose-containing macromolecules on the surface. Their clinical use has since declined, because of their negative contrast *i.e.* reduced MRI signal intensity, and poor biodistribution. The critical role of nanoparticle coatings in biological applications was further highlighted by Wilhelm et al. who conducted a meta-analysis of literature data on solid tumor targeting for nanoparticles of various types, including MNPs.[46] This study revealed that nanomedicine has largely fallen short of early expectations for clinical translation. In animal models, only about 0.7 % of intravenously administered nanoparticles reached the tumor site, with the remainder eliminated by the immune system.[46] For applications in biology, MNPs must be protected from the non-specific adsorption of endogenous biomolecules such as proteins, lipids, amino acids, and nucleotides onto their surface. Extensive studies on the formation of the protein or chemical corona have shown that such adsorption changes the biological identity of the particles,[47] as well as it promotes their aggregation.[48,49] For MNPs, surface coatings serve multiple functions: they modulate interparticle interactions, preserve size and surface properties, and maintain colloidal stability in complex media.[44]

To address the challenges outlined above, numerous state-of-the-art strategies have been developed, and comprehensive reviews have been published in recent years.[4,50-53] A widely used protocol for surface modification involves coating the particles with an organic layer composed of low-molecular-weight ligands, surfactants, phospholipids, or polymers with diverse architectures.[52] This non-covalent modification is typically achieved through a two-step process: first, nanoparticles and coating agents are synthesized independently; then, in a second step, they are combined under suitable physicochemical conditions. The two-step process supports a broad spectrum of surface modification techniques, such as ligand adsorption, polymer grafting, the formation of self-assembled monolayers, and layer-by-layer deposition.[44,50,51,53] A clear example of two-step coating is provided by polymers bearing phosphonic acid functional groups, which attach strongly to iron via multidentate bonds.[53] When combined with an outer poly(ethylene glycol) (PEG) layer, this coating remains highly stable in cellular environments, limiting nonspecific uptake and reducing toxicity.[53-55] Such approaches offer several advantages, including operational simplicity, adaptability, and high yields.

## II.3 - Magnetic properties and forces

Vibrating sample magnetometry (VSM) is the technique dedicated to the magnetic characterization of iron oxide dispersions.[43,56] The experiment consists of measuring the magnetization $M$ *versus* magnetic field $H$ based on Faraday law of induction. An illustration of the $M(H)$ behavior is provided in Fig. 1e for MNPs of diameter 6.2, 8.3 et 11.0 nm. The magnetization curve shows a linear increase at low fields and reaches saturation at high fields. These VSM data are in good agreement with the expected paramagnetism behavior[21]:

$$M(H) = \phi m_S \big(\coth(\xi(H)) - 1/\xi(H)\big) \qquad (2)$$

where $m_S$ is the saturation magnetization, $\xi(H) = \mu_0 m_S v H / k_B T$ is the ratio between magnetic and thermal energy, $\mu_0$ is the permeability in vacuum, $v$ the particle volume, $\phi$ is the MNP volume fraction and $\mathcal{L}(\xi) = \coth(\xi) - 1/\xi$ the Langevin function.[41] Adjustment with experimental data allows the size distribution of magnetic nanocrystals to be determined. The magnetic diameter obtained by VSM is slightly smaller than the structural diameter measured by TEM, which is attributed to the presence of a ~ 1 nm surface layer containing crystalline defects.[33,57] Notably, magnetite and maghemite MNPs exhibit magnetization values between those of metallic iron (Fe⁰) and other oxides such as hematite, making them suitable for applications involving remotely induced motion (Table I).

| Iron oxide nanoparticles | Chemical formula | $\rho$ kg m$^{-3}$ | $m_S$ A m$^{-1}$ |
|---|---|---|---|
| Iron | Fe⁰ | 7874 | 17.8×10⁵ |

| Magnetite‡ | $Fe_3O_4$ | 5170 | $4.4\times10^5$ |
| Maghemite‡ | $\gamma$-$Fe_2O_3$ | 4900 | $3.5\times10^5$ |
| Hematite | $\alpha$-$Fe_2O_3$ | 5300 | $0.21\times10^5$ |

**Table I**: *List of mass densities $\rho$ and volumetric magnetization $m_S$ of iron ($Fe^0$), magnetite ($Fe_3O_4$), maghemite ($\gamma$-$Fe_2O_3$) and hematite ($\alpha$-$Fe_2O_3$). The oxides indexed with a double obelisk are obtained by Massart synthesis.*

The magnetic force exerted on a single nanoparticle by a magnetic field gradient $\boldsymbol{\nabla} H$ expresses as $\boldsymbol{F}_{NP}(H) = \mu_0(\boldsymbol{m}\cdot\boldsymbol{\nabla})H$, where $m$ its magnetic moment. For a field gradient along the *z*-direction, one gets:

$$F_{NP}(H) = \mu_0 m_S v \mathcal{L}(\xi(H)) dH/dz \qquad (3)$$

For a dispersion of 10 nm MNPs placed in a magnetic field generated by a bench-top neodymium magnet with $H = 2\times10^5$ A m$^{-1}$ and $dH/dz = 2\times10^7$ A m$^{-2}$, this force is of the order of $10^{-18}$ N, and therefore very small. In the magnetophoretic experiment shown in Fig. 1f, applying such a field to a dilute MNP dispersion results in an exponential concentration gradient, with the highest concentration near the magnet.[33] However, concentration redistribution is slow, typically taking several days to reach equilibrium. This delay in reaching a steady state is due to the fact that for sub-10 nm MNPs the magnetic energy is lower than the thermal energy, i.e. $\xi < 1$. To achieve remote manipulation on shorter time scales, MNPs must be encapsulated in micron-sized assemblies with high iron oxide loadings. Figs. 1g and 1h show aggregates of Massart MNPs formed by electrostatic complexation of anionic particles with cationic polymers. Depending on the complexation scheme, these aggregates may contain from a few dozen nanoparticles (Fig. 1g) to several thousands (Fig. 1h), greatly enhancing magnetic separation under an applied field. Finally, it is worth noting that MNPs can also spontaneously assemble into chains several hundred nm long, as illustrated in Fig. 1i.[58-60] This chaining results from magnetic dipolar interactions, which become significant for particles with sizes larger than 20 nm. In the latter case, the particles are not held together by physical bonds, but by electromagnetic forces.

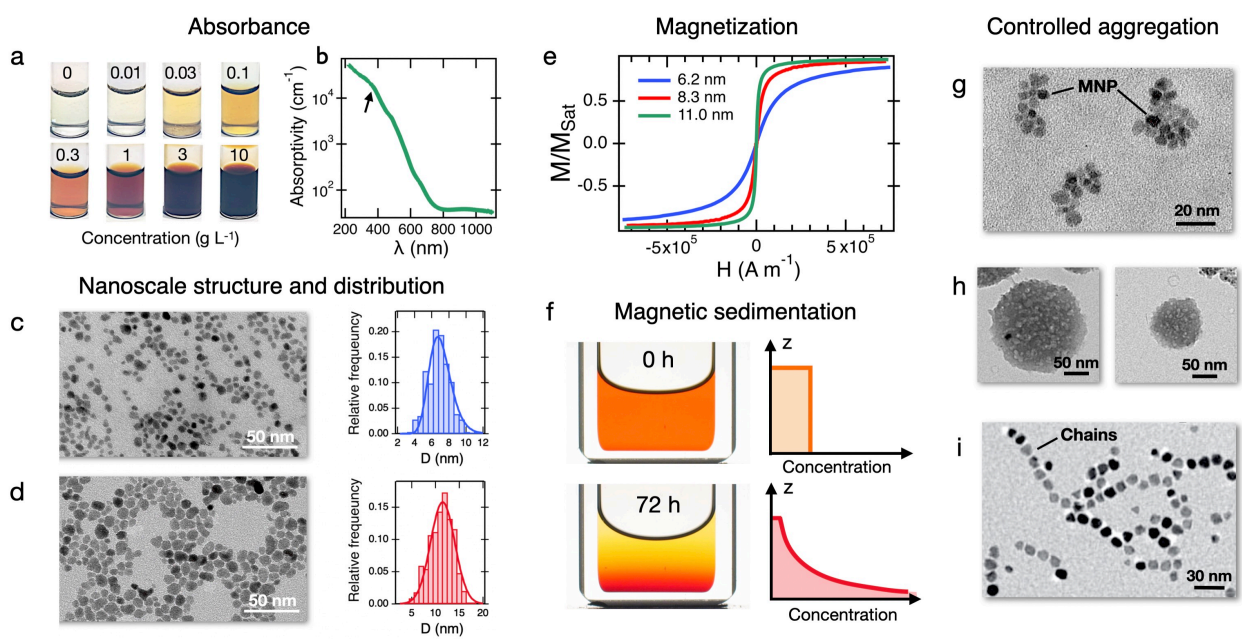

*Figure 1: a)* Dispersions of maghemite (γ-Fe$_2$O$_3$) at concentrations ranging from 0 (pure water) to 10 g L$^{-1}$. With increasing concentration, the color evolves from yellow to orange and finally to garnet red at the highest concentration. *b)* Mass absorptivity of a γ-Fe$_2$O$_3$ dispersion as a function of wavelength. Data correspond to an 8.3 nm maghemite batch, with concentration independently determined by inductively coupled plasma mass spectrometry.[61] This calibration curve is routinely used to quantify dispersion concentrations. *c–d)* TEM images of γ-Fe$_2$O$_3$ nanoparticles with median sizes of 6.2 and 11.0 nm, together with the corresponding size distributions (right panels). *e)* Magnetization curves of γ-Fe$_2$O$_3$ dispersions with particle sizes of 6.2, 8.3, and 11.0 nm, normalized to their saturation values. *f)* Photographs (left) and concentration profiles (right) of a magnetic dispersion at the initial (t = 0 h) and final (t = 72 h) stages of magnetic sedimentation. The magnet is placed at the bottom of the cell, generating a constant magnetic field gradient over the sample height (adapted from Ref.[33]). *g,h)* MNP aggregates obtained by electrostatic complexation using direct mixing with amphiphilic block-copolymers[62] or dialysis method,[63] respectively. *i)* In situ cryo-TEM images of vitrified magnetite dispersions containing 24 nm iron oxide nanoparticles. Magnetic dipolar interactions lead to anisotropic assemblies including flexible chains, branched chains, and flux-closure rings (adapted from Ref.[60]). The particles shown in this panel were not produced using the Massart synthesis method.

# III – Nanoparticles and macroscopic forces on cells and tissues

## III.1 – Massart magnetic nanoparticle–cell interaction, biocompatibility and toxicity

In this section, we focus on the interactions between MNPs and living cells under standard 2D monolayer culture conditions. The behavior of MNPs in contact with cells has been the subject of intensive investigation over the past two decades. The aim here is to outline the key principles that explain how MNP-cell interactions unable their applications in cellular and tissue mechanics. For broader perspectives on pharmaceutical applications, systemic administration, or cell labeling, the reader is referred to existing reviews.[17,23]

An important requirement for controlling MNP internalization in cells is functionalization, that is, the addition of a coating layer at the particle surface. Without this protecting layer, the MNPs precipitate in the culture medium due to the pH difference between synthesis and physiological conditions, rapidly forming micrometer-sized aggregates.[37,49] Among the coatings designed for MNPs, ligands and polymers are the most frequent, with several well-established examples including citrate ions, PEG, poly(vinyl alcohol), poly(ethylene imine), poly(acrylic acid) (PAA), chitosan, dextran, or alginate.[17] Short polymer PEGylated chains, combining iron-specific anchoring moieties, form surface brushes with a thickness of 5 to 10 nm that increase protein resistance and improve biocompatibility.[53]

Table II summarizes the cell types tested with Massart nanoparticles, illustrating their broad applicability in biology. In these examples, MNPs are internalized *via* endocytosis, a natural process in which the plasma membrane engulfs extracellular material to form vesicles. There, nanoparticles are confined within endosomal compartments of 200–500 nm, without direct contact with the cytosol.[64] This compartmentalization prevents interactions with intracellular machinery and supports overall biocompatibility. The MNP loaded endosomes remain stable for several days, allowing magnetically labeled cells to be tracked and manipulated later on. Examples of such intracellular endosomes are shown in the TEM images in Figs. 2a and 2b.

| Cell type | Example cell line(s) | Typical application with MNPs | References |
|---|---|---|---|
| Immune cells | RAW 264.7 macrophages | Uptake studies, immunological response | 54,65 |
| Endothelial cells | HUVEC | Endocytosis, transport | 66 |
| Epithelial cells | MCF-10A, MDCK | Endocytosis, transport | 39 |
| Connective tissue cells | Chondrocytes, NIH/3T3 fibroblasts | ECM production, mechanobiology | 67,68 |
| Muscle cells | C2C12 | Differentiation, force generation | 69 |
| Stem cells | Primary cells, MSCs | Labeling, differentiation, regenerative medicine | 70-72 |
| Neuronal models | PC12 | Neurobiology, axonal growth | 73 |
| Cancer cells | HeLa, MCF-7, A549 | Uptake, drug delivery, mechanobiology | 38,39 |

***Table II:*** *Representative cell types studied with Massart magnetic nanoparticles. The table lists example cell lines, typical applications (e.g. uptake studies, mechanobiology, tissue engineering), and selected references using nanoparticles synthesized by the Massart coprecipitation method.*

Observations of MNP localization are complemented by quantitative measurements of internalized iron using techniques such as inductively coupled plasma spectroscopy, magnetophoresis, relaxometry and colorimetric assays.[68,74-76] Results are generally reported in picograms of iron per cell, with typical values ranging from $10^{-2}$ to 100 pg for administered doses between 1 and 1000 μg mL$^{-1}$. The main parameters governing MNP uptake are the surface coating, the applied dose, and the incubation time.[54,68,77,78] For instance, 8.3 nm particles uptaken by NIH/3T3 murine fibroblasts at a concentration of 800 μg mL$^{-1}$ yielded internalized amounts of 0.1, 10, and 100 pg Fe per cell when coated with hydrophilic copolymers, PAA$_{2k}$ and citrate ligands, respectively, highlighting the critical role of surface chemistry in regulating iron uptake.[68] Iron doses in the range of 0.5–5 pg per cell have been shown to promote aggregate formation in response to a magnetic field without inducing cytotoxicity or altering differentiation and cell fate.[78] Over periods longer than a week, MNPs may degrade as endosomes fuse with lysosomes, where acidic conditions trigger their dissolution. This process releases iron ions that can generate reactive oxygen species via the Fenton reaction.[79] However, cells mitigate these effects through iron homeostasis, notably ferritin storage and recycling into hemoglobin.[80] At sufficiently low nanoparticle concentrations, these mechanisms neutralize toxicity, as observed in several cell lines.[77]

### III.2 - From 2D cell culture to model micro-tissues and organoids

*III.3.1 - Motivation for 3D magnetic cell assembly*

Studies have shown that traditional 2D cultures on plastic or glass substrates fail to mimic the natural cell environment in terms of stiffness and topography.[81,82] By contrast, 3D cultures more closely reproduce *in vivo* conditions, and mimic the influence of neighboring cells and extracellular matrix on key processes such as migration, differentiation, morphogenesis, and gene expression.[83,84] In tumors, microenvironmental alterations can drive the transition from a non-invasive to an invasive state and promote epithelial–mesenchymal transition.[85,86] Such reciprocal interactions where the microenvironment influences cell behavior and, in turn, cells remodel their microenvironment, can only be captured in 3D cell cultures.[87,88]

Over the past decade, two principal strategies have been developed for constructing model 3D tissues: one uses exogenous matrices, designed through materials chemistry from synthetic polymers or purified proteins to recreate an extracellular environment, while the other relies on engineering cell aggregation and exploits the ECM secreted naturally by the cells. The challenge with the first strategy lies in replicating the architecture, roughness and composition of native tissues, which must be tailored to each specific tissue type.[89,90] The second strategy, also known as the scaffold-free approach, involves creating multicellular assemblies that build on the self-organization of cells and ECM-cell feedback interactions to establish tissue-like structures. 3D cell culture provides a high degree of clinical and biological relevance to *in vitro* models, especially spheroid cultures.[91] The main technological challenge of the scaffold-free approach is achieving cell assemblies with well-controlled geometry and composition, a task in which magnetic nanoparticles can play a crucial role.

Scaffold-free microtissue formation typically occurs in two stages, the aggregation of individual cells into a compact assembly, and the maturation of cell–cell contacts that confer the tissue its integrity. In this case, the aggregation stage governs the overall morphology and architecture of the MNP assembly. Without external forces, cell aggregation is slow and may require up to a week.[91] The aggregation step can be accelerated by applying external forces, such as gravity,[92] centrifugation[93] random agitation on cell-repellent substrates,[94] or through encapsulation methods.[95] These strategies generally require one to three days to produce aggregates of a few hundred micrometer, comprising approximately $10^5$-$10^6$ cells. By contrast, magnetic forces accelerate the process further, generating millimeter-sized aggregates within only a few hours.[96]

*III.3.2 – Forces applied to magnetically labelled cells*

For remote manipulation, adherent cells in 2D culture must first be labeled with MNPs. As noted earlier, the particles are spontaneously internalized by endocytosis and accumulate in the cytoplasm within endosomes (Fig. 2). Under a magnetic field gradient, each of these compartments experiences a magnetic force proportional to the number $n$ of enclosed particles. Assuming in addition an average of $N$ labeled endosomes per cell, the total force exerted on a single cell becomes:

$$F_{Cell}(H) = nNF_{NP}(H) \qquad (4)$$

By estimating a magnetic nanoparticle volume fraction of 10 percent within each endosome and an average of 100 iron-loaded endosomes per cell,[97] the resulting magnetic forces are on the order of $10^{-2}$ pN per endosome and approximately 1 pN per cell (Table III). In a culture medium with a viscosity of $10^{-3}$ Pa s, once the cells are detached from their substrate, this force induces a linear displacement with a velocity given by Stokes law, which in this case is about 100 µm s$^{-1}$, thereby driving their accumulation at the magnet interface (Fig. 2c). These estimates show that, at non-toxic iron doses, magnetic forces can drive cell magnetophoresis over centimeter-scale distances within a few minutes.

| | $D$ (nm) | Volume fraction $\phi$ (%) | MNP number $n, N$ | Mass of iron (pg) | Magnetic Force (N) |
|---|---|---|---|---|---|
| **10 nm diameter nanoparticle** | 10 | 100 | 1 | $1.8 \times 10^{-6}$ | $2.7 \times 10^{-18}$ |
| **Endosome** | 500 | 10 | 12500 | 0.022 | $3.4 \times 10^{-14}$ |
| **Living cell** | 10000 | 0.1 | $10^6$ | 1.8 | $2.7 \times 10^{-12}$ |

***Table III:*** *Characteristic magnetic forces acting on single MNP compared to those on magnetically labelled endosomes and living cells. Here $D$ denotes the diameter and $\phi$ the MNP volume fraction.*

*The mass of iron internalized is given in picogram. For the evaluation of the magnetic force, it is assumed (as in Section II) that the field is generated by bench-top neodymium permanent magnets with $H = 2\times10^5$ A m$^{-1}$ and $dH/dz = 2\times10^7$ A m$^{-2}$. Values for the endosomes and cells are given as typical orders of magnitude reported in the literature. Note that magnetic forces are weaker than the adhesive forces typical for adherent cells (1 nN)[34] and thus cannot detach them, restricting this approach to cells in suspension or previously trypsinated.*

*III.3.3 - Magnetic strategies for 3D cell assembly*

The scaffold-free approach has given rise to three main protocols: magnetic levitation of cell sheets,[98,99] magnetic patterning,[96,100] and magnetic molding.[70,101] In the first approach, an external magnetic field drives cells in suspension to the air–liquid interface, where they aggregate over a few days to form three-dimensional patches.[99,102] It is applicable to numerous cell types, including stem cells and co-cultures,[103] but its main limitation lies in the poor control of assembly architecture due to weak lateral constraints at the air–liquid interface. In the second method, a magnet is placed beneath a Petri dish containing magnetically labeled cells in suspension.[71,104,105] The cells are attracted toward the substrate and spontaneously arrange into patterns that reflect the shape of the underlying magnet (Figs. 2d-f). Main advantages here are the versatility of aggregate geometries, reproducible formation, and straightforward implementation. Magnetic molding is a more recent technique that employs agarose molds shaped with steel spheres, cylinders, or cubes held in place by magnets during the agarose gelation (Figs. 2g-i). Once solidified, these molds serve as wells into which magnetically labeled cells are deposited, again guided by magnetophoresis.[70] After the magnet is removed, tissue maturation takes place, leading to a cohesive structure typically within 12 hours. This method enables rapid and reproducible production of aggregates with sizes and shapes not achievable otherwise.[69] All three strategies have been demonstrated using Massart magnetic nanoparticles, with applications spanning drug testing, cancer therapy, tissue engineering, mechanobiology, and bioprinting.[100]

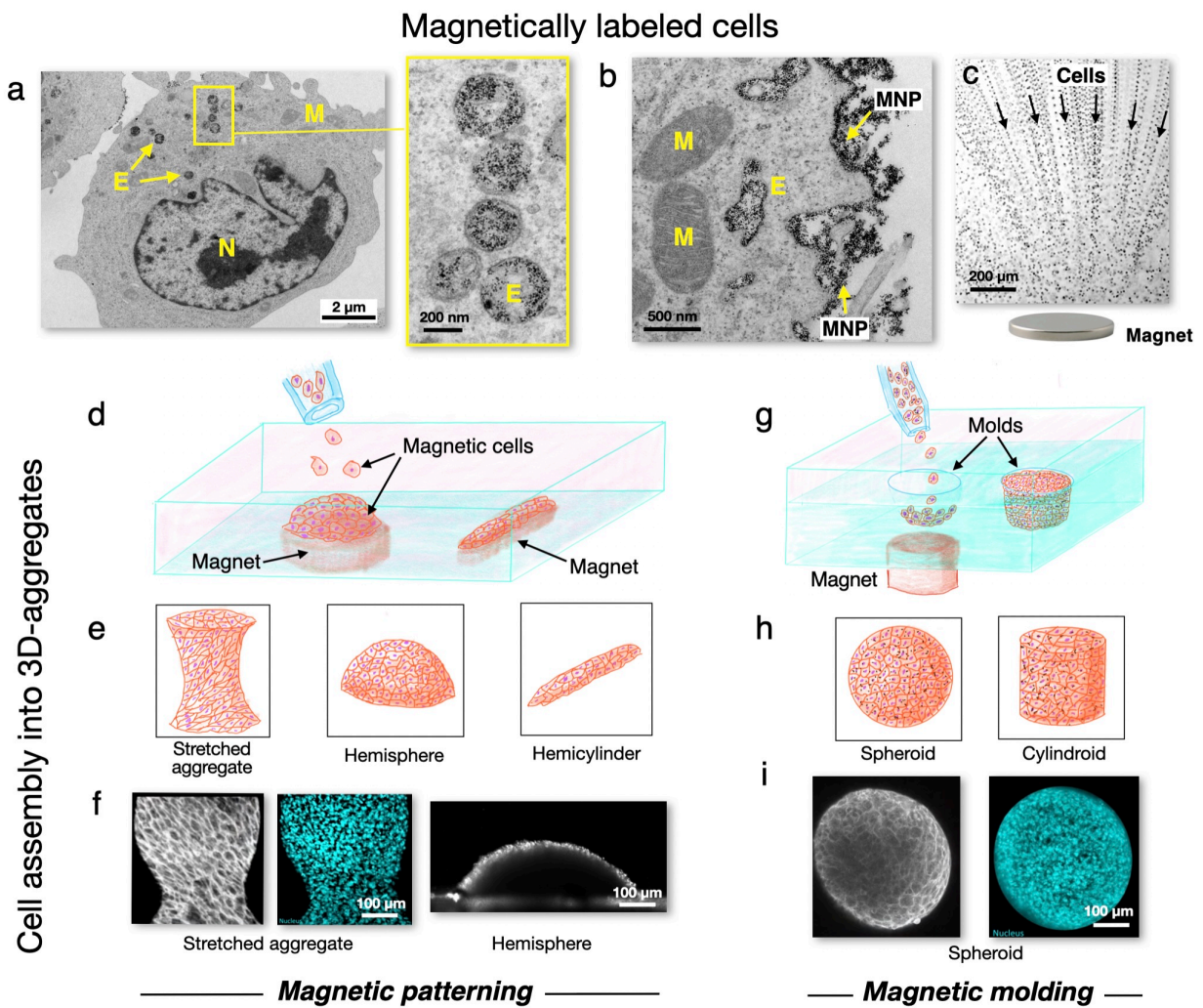

*Figure 2: a)* TEM images of a cell incubated with 800 µg mL$^{-1}$ iron oxide nanoparticles (Massart synthesis, 11 nm, citrate-coated), with incubation time of 4 hours. Inset: magnified view showing five endosomes containing MNPs.[37] The letters E, M and N denote the endosomes, mitochondria and nucleus, respectively. *b)* Same conditions as in (a), but after 10 minutes of incubation. The image shows precipitation of citrate-coated MNPs at the plasma membrane, with deposits up to 500 nm thick.[68] An invagination of the membrane near a newly formed endosome is also visible (arrow).[37] *c)* Kymograph of magnetic cells attracted toward a magnetic microtip. Each dot represents a single cell, illustrating the magnetic field lines. The pattern reflects the strength and gradient of the magnetic forces experienced by the cells (adapted from Ref.[71]). *d)* Schematic of magnetic patterning for scaffold-free cell aggregate formation. Magnetically labeled cells are drawn toward a magnet beneath the Petri dish, spontaneously assembling into hemi-spherical or cylindrical aggregates[71,104] *e)* Examples of aggregate morphologies obtained by magnetic patterning. From left to right in panel: a stretched aggregate,[106] a hemisphere, and a hemicylinder. *f)* Experimental formation of C2C12 myoblast aggregates: the first two images show confocal microscopy with actin and nucleus labeling, respectively, and the third image shows a truncated hemisphere observed by bright-field microscopy. *g)* As for d) for magnetic molding. This technique uses agarose molds shaped with steel objects held by magnets, creating wells into which cells are *assembled.*[70] *h) Examples of aggregate morphologies obtained by magnetic molding: from left to right, a spheroid and a cylindroid.*[106] *i) C2C12 myoblast aggregates imaged by confocal microscopy with actin and nucleus staining, corresponding to the morphologies shown in h). Figs. 2f and 2i are adapted from Ref.*[106]*.*

## III.3 - Applying magnetic forces to cell aggregates: towards rheology of biological tissues

*III.3.1 - From local to macroscopic deformations*

Biological tissues are soft, complex, and heterogeneous materials composed of multiple cell types, whose collective behavior often differs from that of the same cells studied individually. Their rheological properties can be probed using either endogenous or exogenous mechanical actuators. Among the latter, established methods include conventional rheology[107] performed in shear or compression–extension modes,[94,108] AFM,[109] micropipette aspiration,[110,111] and microfluidics.[112,113] Because direct mechanical contact may induce structural reorganization, less invasive approaches have been devised, relying on embedded sensors or on external stimuli.[114-116] Magnetic actuation is particularly relevant in this context, as the labeled cells themselves serve as vectors for transmitting mechanical stresses throughout the entire tissue.[117] Typical experimental conditions involve fields of $1-5\times10^5$ A m$^{-1}$ and gradients $1-5\times10^7$ A m$^{-2}$. With cell densities around $10^6$ cells per mm³, the resulting volumetric forces range from $10^{-6}$ to $10^{-4}$ N per mm³ (see Table IV).[70,108] These forces tend to flatten multicellular aggregates and oppose intrinsic tissue forces such as surface tension,[36,69] cell–cell adhesion,[26] and actomyosin contractility.[35] MNPs have also been used *in vivo* in zebrafish,[118] microalgae, and echinoderms,[119] to reproduce morphogenetic movements,[100,118,120] as well as tumor growth,[121] or to simulate the mechanical effects of stent implantation.[122]

| Type of force | Context | Typical magnitude | Reference |
|---|---|---|---|
| **Cell–cell adhesion forces** | Adherens junctions in epithelia / tissues | 10–100 nN per junction | 26 |
| **Cell–substrate traction** | Focal adhesions, fibroblasts on ECM | 10–100 nN per adhesion site; up to ~1 µN per cell | 34 |
| **Tissue contractility** | Collective forces in sheets and tissues | µN per mm² | 35 |
| **Tissue Surface tension** | Surface forces in spheroids | 1-10 mN m$^{-1}$ | 36,69 |
| **Magnetic forces on MNP-loaded cells** | Endocytosed Massart nanoparticles in field gradients | 10–100 pN per cell (moderate gradients), up to nN per cell (strong gradients) | 123 |

*Table IV: Characteristic forces generated by cells and their microenvironment, such as cell-cell adhesion, cell-substrate traction, tissue contractility, and tissue surface tension, compared with magnetic forces acting on cells loaded with Massart magnetic nanoparticles (MNPs). Magnetic forces are generally several orders of magnitude lower than intrinsic cellular and tissue forces, ranging from tens of picoNewtons to nanoNewtons depending on field strength.*

## III.3.2 - Viscoelastic Behavior of Multicellular Aggregates

Here, we present examples of mechanical measurements on multicellular aggregates controlled by magnetic fields. Figs. 3a and 3b show a spheroid of C2C12 muscle cells before and after exposure to a permanent magnet until steady state is reached. With the magnet placed underneath, the spheroid flattened by about 10%. From the spheroid profiles with and without the applied force, a surface tension of 21 mN m$^{-1}$ and a Young modulus of 100 Pa are estimated using Laplace and Hertz laws.[69] A second illustrative experiment on reconstructed tissues is the compression creep test performed on a cylindrical aggregate of F9 mouse embryonal cells (Figs. 3c and 3d). In this setup, the electromagnet enables experiments either at constant force, or at varying frequencies between 10$^{-2}$ and 100 rad s$^{-1}$. Fig. 3e shows a kymograph of the upper surface of an aggregate subjected to a 10 s magnetic stimulation and the subsequent release of the field. The corresponding deformation $\varepsilon(t)$ is initially rapid (< 1 s) and mainly elastic, followed by a creep process reflecting cellular reorganization.[70,94,115] Upon field release, the aggregate height recovers and, over sufficient time, returns to its original value (Fig. 3f). This experiment highlights the viscoelastic nature of multicellular aggregates, with an elastic response, a creep regime, and a reversible recovery.[107] The insert in Fig. 3f shows the compliance $J(t)$ plotted *versus* time, revealing a power-law behavior of the form $J(t) \sim (t/\tau)^\alpha$ with an exponent $\alpha$ of 0.24. This behavior is characteristic of a fractional viscoelastic model, consistent with a wide distribution of relaxation times in the cellular aggregate.[124,125] Measurements on cell aggregates have consistently yielded viscoelastic behavior with Young moduli in the range of 10–100 Pa, which is 1–2 orders of magnitude lower than that of the individual cells composing the aggregates.[69] This apparent softening of aggregates reflects the collective effects of cell–cell junctions and cellular rearrangements.

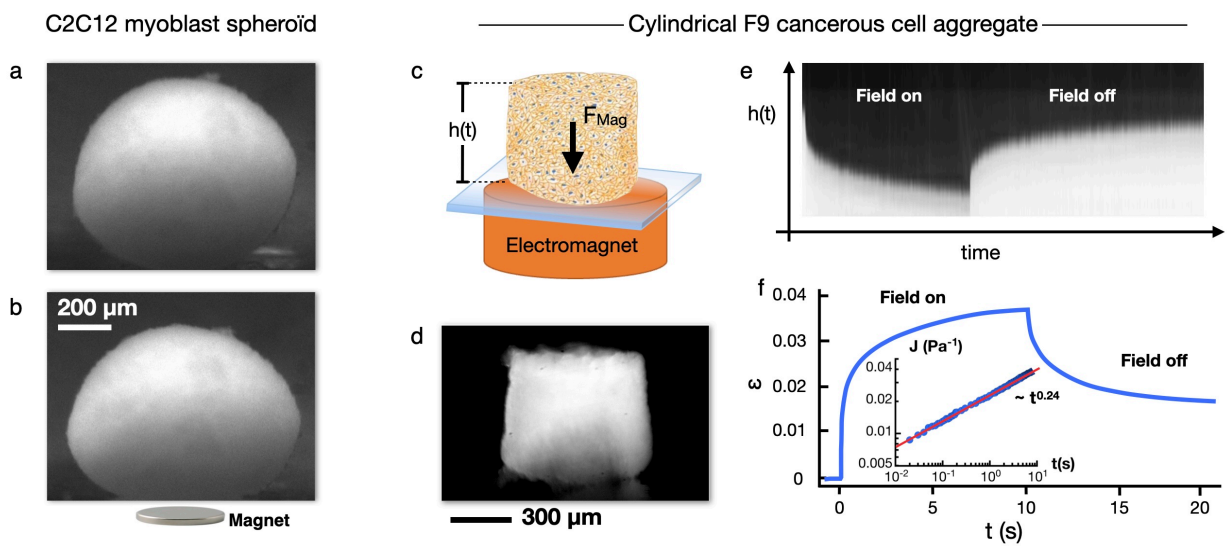

*Figure 3: a,b) Representative profiles of a C2C12 spheroid. The top image shows a spheroid before magnet application, and the bottom image after magnetic flattening at mechanical equilibrium (adapted from Ref.[69]). c) Schematic of a cylindrical multicellular aggregate obtained by magnetic molding (Fig. 2g) and subjected to the field of an electromagnet. In this setup, the cells can experience either constant forces or frequency-dependent forces, mimicking macroscopic rheology experiments. d) Cylindrical aggregate of F9 murine embryonal cells.[70,71] e) Kymograph of the upper surface of the aggregate shown in (d) during a 10-second step of applied magnetic force. f) Deformation $\varepsilon(t)$ of the same aggregate during an on-off magnetic field sequence, illustrating the mechanical response of the tissue. Inset: compression compliance J(t) as a function of time, calculated from $\varepsilon(t)$, displaying a power-law behavior characteristic of a fractional viscoelastic model.[124,125]*

# IV – Magnetic wires for microrheology and mechanobiology

**IV.1 – Magnetic wires for active microrheology**

*Electrodeposition in nanoporous templates*

Micron-sized anisotropic objects, whether ferromagnetic or paramagnetic, respond not only to magnetic forces but also to magnetic torques. The rotational dynamics of a magnetic wire immersed in a medium provide direct access to local rheological properties, enabling experimental determination of viscosity and elasticity. The first significant attempt in this direction was reported by Crick and Hughes,[126] who introduced anisotropic particles into chick fibroblast cells and analyzed their orientational recoil after applying an external magnetic field. Through this experiment, they were able to demonstrate the viscoelastic nature of the cytoplasm. After several decades of limited progress, the development of nanotechnologies in the early 2000s enabled the fabrication of metallic wires, also referred to as needles or rods, and often with the prefix *nano*. These wires were synthesized by electrodepositing metallic materials, e.g. nickel, cobalt, or iron within anodized aluminum oxide membranes containing an array of cylindrical nanopores.[3,127] The nanopores there serve as templates, guiding the growth of the wires with controlled diameter (< 1 µm) and length (10-100 µm). Examples of such metallic wires are displayed in Fig. 4a-b.[128-130] Metal nanowires are ferromagnetic, with saturation magnetization values that exceed those of iron oxide nanoparticles, allowing the wires to generate significant torques. This fabrication strategy has stimulated extensive research in micromechanics,[131-141] particularly in contexts where only small volumes of fluid are available,[142,143] or in biological fluids.[144,145]

*Dialysis-induced magnetic nanoparticle assembly*

In contrast to template-assisted electrodeposition, this approach relies on the spontaneous one-dimensional assembly of Massart MNPs in solutions.[62,63,146] The primary building blocks are maghemite nanoparticles, which are precoated with $PAA_{2k}$ to ensure colloidal stability and confer a high surface charge.[146] Acid–base titration of $MNP@PAA_{2k}$ particles reveals a surface charge density of approximately −20*e* nm$^{-2}$, significantly higher than that of uncoated nanoparticles, with a polymer layer thickness of 2 to 3 nm.[57] The anionic MNP@PAA dispersion is then mixed with a cationic polyelectrolyte, e.g. poly(diallyldimethylammonium chloride), in a high-salt medium. Under such conditions, electrostatic interactions are screened, and no aggregation occurs. Gradual removal of salt by dialysis progressively restores electrostatic attraction below a critical ionic strength, triggering the co-assembly between the oppositely charged components. In the absence of an external magnetic field, the process yields spherical aggregates with diameters of approximately 200 nm,[62,63] in which the particles are held together by the polycation binder (Fig. 1h). When dialysis is performed under a magnetic field, aggregates align along the field direction, ultimately leading to magnetic wires with lengths between 1 to 500 μm and diameters between 0.2 and 2 μm.[63,146] Images of magnetic wires, obtained using optical microscopy, TEM, scanning electron microscopy (SEM), and X-ray energy-dispersive spectroscopy (EDX) are shown in Figs. 4c–g. Small-angle X-ray scattering measurements reveal a short interparticle spacing of about 0.5-1 nm, corresponding to a volume fraction of 30 vol. % for the wire[62] Mechanical characterization, performed *via* magnetic bending experiments, yields a Young modulus on the order of 10 MPa.[147] This indicates that the wires behave as rigid bodies, showing negligible deformation when put under rotation in a medium. The robustness of this mechanism has been demonstrated with diverse MNP/polymer systems by varying the crosslinker chemistry, adjusting the particle size, or incorporating quantum dots to achieve dual magnetic–fluorescent functionality.[148,149]

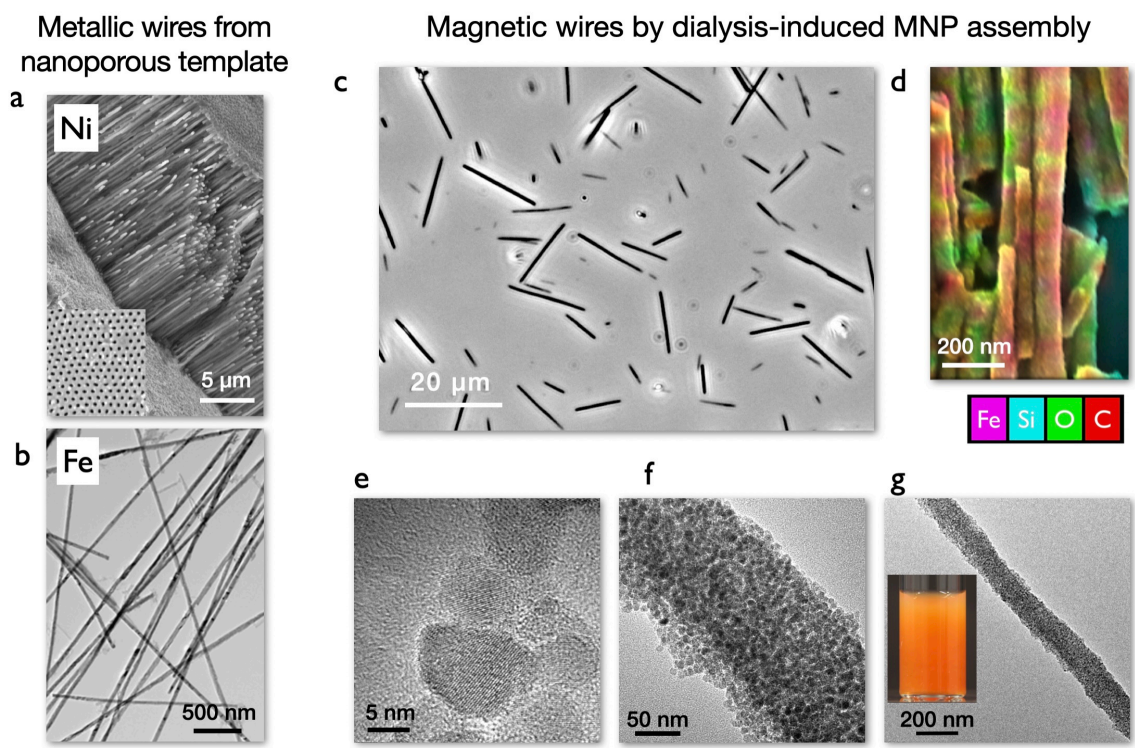

***Figure 4: a)*** *Field-emission scanning electron micrograph of a nickel nanowire network integrated within an anodic aluminum oxide membrane featuring ordered channels with 200 nm pores. Inset: top-view SEM*

*image of the aluminum oxide template prior to electrodeposition (adapted from Ref.[128]). **b)** TEM image of 50 nm-thick iron nanowires fabricated by electrodeposition within the nanopores of an aluminum oxide template (adapted from Ref.[130]). **c)** Phase-contrast optical microscopy image of magnetic wires obtained by dialysis of Massart iron oxide nanoparticles. **d)** Scanning electron microscopy image of magnetic wires superimposed to EDX elemental maps of iron, oxygen, carbon, and silicon of sonicated wires. **e–g)** TEM images of a magnetic wire at different magnifications, showing the crystallinity of $\gamma$-$Fe_2O_3$ (maghemite) nanocrystals and their cylindrical organization. Inset: vial containing a dispersion of magnetic wires.*

**IV.2 – Magnetic Rotational Spectroscopy (MRS) in complex fluids**

MRS is based on applying a rotating magnetic field and tracking magnetic wire motion by optical microscopy varying parameters such as temperature, driving frequency, or wire length.[134,135,142,150,151] The term Magnetic Rotational Spectroscopy originates from mechanical systems whose response displays a resonance peak at a characteristic frequency.[152] There are three canonical models for describing the linear rheological properties of simple and complex fluids: the Newtonian fluid, represented by a dashpot; the Maxwell fluid, described by a dashpot and spring in series (Fig. 5a); and the Standard Linear Solid (SLS) model, consisting of a Maxwell element in parallel with an additional spring (Fig. 5b). Within the framework of linear rheology complex fluids can be described as a suitable combination of these elementary models. The constitutive equations governing wire dynamics admit solutions that provide the framework for interpreting MRS experiments.

*IV.2.1 – Newtonian fluid*

A Newtonian fluid is characterized by a finite zero-shear shear viscosity, denoted $\eta_0$. At sufficiently low frequencies, the wire rotates in phase with the applied field, exhibiting synchronous motion. As the frequency increases, the viscous torque grows, and beyond a critical frequency $\omega_C$, the wire can no longer follow the field. The transition between synchronous and asynchronous rotation thus occurs at this critical frequency:[150,151]

$$\omega_C = \frac{3}{8\mu_0} \frac{\Delta\chi}{\eta_0} g\left(\frac{L}{D}\right) \frac{D^2}{L^2} B^2 \qquad (5),$$

where $L$ and $D$ denote the length and diameter of the wire, $\Delta\chi$ the anisotropy of susceptibility, and $g(x) = ln(x) - 0.662 + 0.917x - 0.050x^2$.[38] Since $\omega_C \sim 1/\eta_0$ in Eq. 5, viscosity can be directly obtained from the measurement of the critical frequency. Two quantities related to the wire motion can be extracted from the wire motion: the average rotation velocity $\Omega(\omega)$ and the amplitude of the back-and-forth oscillations $\theta_B(\omega)$. For a Newtonian fluid, $\Omega(\omega)$ exhibits a cusp-shaped maximum, similar to a resonance centered on $\omega_C$, with its frequency dependence described by:

$$\omega \leq \omega_C \qquad \Omega(\omega) = \omega$$

$$\omega \geq \omega_C \qquad \Omega(\omega) = \omega - \sqrt{\omega^2 - \omega_C^2} \tag{6}$$

In a purely viscous liquid, for $\omega > \omega_C$, $\theta_B(\omega)$ decreases as $1/\omega$ and therefore tends towards zero at high frequencies.[153,154] The predictions of Eqs. 5 and 6 have been experimentally observed in a variety of Newtonian fluids, including viscosity standards,[63,129,150] dilute polymer solutions,[154] thin films of ceramic precursors,[143] and biological fluids.[155,156]

*IV.2.2 – Maxwell viscoelastic fluid*
For a Maxwell fluid, $\eta_0$ is expressed as the product of the shear modulus $G_0$ and $\tau$ the relaxation time which characterizes the time-dependent viscoelastic response. As in a Newtonian fluid, a synchronous–asynchronous transition is observed, and Eq. 5 remains valid, provided that $\eta_0$ is replaced by $G_0\tau$ in the expression (Fig. 5c and 5d). Similarly, $\Omega(\omega)$ exhibits a resonance peak that corresponds to the behavior described by Eq. 6 (Fig. 5e). The only difference between Newtonian and Maxwell fluids lies in the asynchronous regime: the oscillation amplitude $\theta_B(\omega)$ remains finite at high frequencies and varies inversely with the elastic modulus (Fig. 5f):

$$\lim_{\omega \to \infty} \theta_B(\omega) = \theta_0 = \frac{3}{4\mu_0} \frac{\Delta\chi}{G_0} g\left(\frac{L}{D}\right) \frac{D^2}{L^2} B^2 \tag{7}$$

Of particular interest is the relationship between Eqs. 5 and 7, $2\omega_C \tau = \theta_0$. Again, the predictions of Maxwell model were fully confirmed experimentally, using e.g. semi-dilute wormlike micellar surfactant solutions.[153]

*IV.2.3 – Standard Linear Solid model*
Soft solids are characterized by a yield stress $\sigma_Y$, meaning that they flow only when the applied stress exceeds $\sigma_Y$. Their behavior can be described by the SLS model.[125,157] For stresses below $\sigma_Y$, the material undergoes only elastic deformation. In this regime, the critical frequency vanishes ($\omega_C = 0$, Figs. 5g and 5h), and the system exhibits only asynchronous oscillations, whose amplitudes are related to the elastic moduli $G_{eq}$ and $G_0$ through the expressions:

$$\lim_{\omega \to 0} \theta_B(\omega) = \theta_{eq} \text{ and } \lim_{\omega \to \infty} \theta_B(\omega) = \frac{\theta_0 \theta_{eq}}{\theta_0 + \theta_{eq}} \tag{8}$$

$$\text{where } \theta_{eq} = 3\Delta\chi B^2/4\mu_0 G_{eq} L^{*2} \text{ and } \theta_0 = 3\Delta\chi B^2/4\mu_0 G_0 L^{*2} \tag{9}$$

where $G_{eq}$ is the equilibrium modulus (Fig. 5i). The SLS model predictions were also confirmed using bacterial hydrogels made from calcium crosslinked polysaccharides.[154] As shown in Table V, MRS readily differentiates a Newtonian liquid, a viscoelastic fluid, and a soft solid. The quantities retrieved from active wire microrheology, $\omega_C$, $\Omega(\omega)$ and $\theta_B(\omega)$ exhibit characteristic frequency dependencies that allow the rheological nature of the fluid to be identified unambiguously. Finally, the MRS technique with Massart nanoparticles can probe viscosities from $10^{-3}$ to 1000 Pa s and moduli from 0.05 to 500 Pa, thus covering one of the widest ranges of simple and complex fluids, both synthetic and biological.

| Rheological models | Static shear viscosity | Yield stress | Sync./Async. transition | $\Omega(\omega)$ | $\lim_{\omega \to \infty} \theta_B(\omega)$ |
|---|---|---|---|---|---|
| **Newton** | $\eta_0$ | No | Yes | Eq. 6 | 0 |
| **Maxwell** | $\eta_0 = G_0 \tau$ | No | Yes | Eq. 6 | $\theta_0$ |
| **Standard Linear Solid** | infinite | Yes | No | 0 | $\theta_0 \theta_{eq} / (\theta_0 + \theta_{eq})$ |

***Table V***: *List of models used in rheology to describe the linear mechanical response of fluids and soft materials, along with predictions for wire rotation in synchronous and asynchronous regimes. In column 2, $\eta_0$ denotes the zero-shear viscosity, $G_0$ the instantaneous elastic modulus, and $\tau$ the relaxation time. For notations specific to MRS, $\Omega(\omega)$ represents the average rotation velocity, while $\theta_{eq}$ and $\theta_0$ denote the oscillation angles at low and high angular frequency, respectively.*

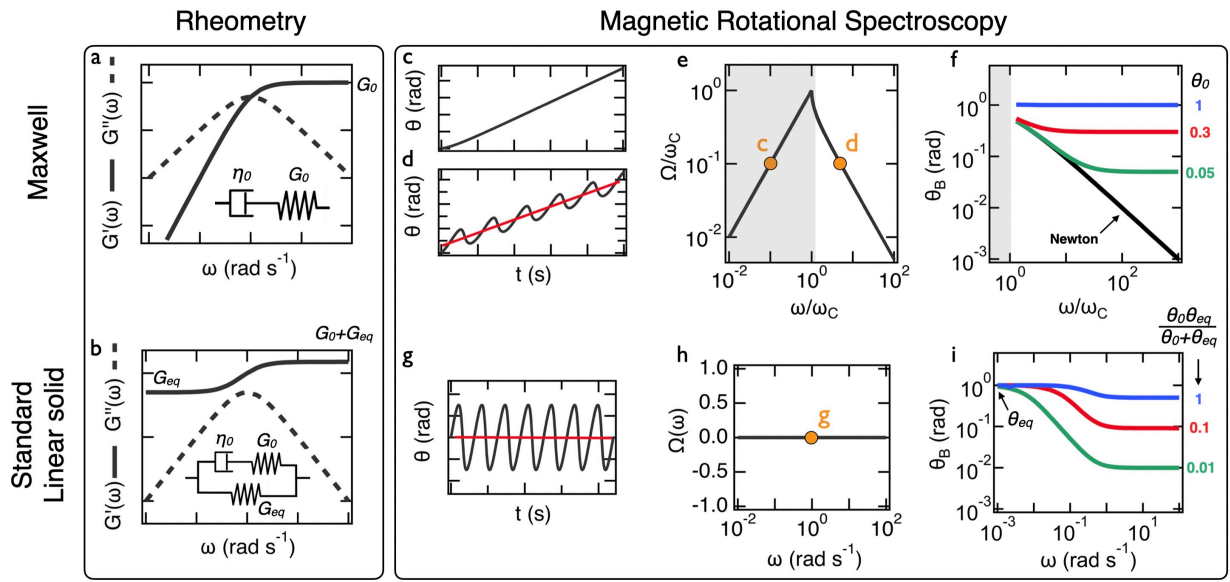

***Figure 5: a,b)*** *Schematic representation of the storage (G') and loss (G") moduli as functions of the applied angular frequency, according to the Maxwell and Standard Linear Solid models. Insets illustrate the corresponding mechanical analogs composed of viscous dashpots and elastic springs.* ***c,d)*** *Orientation angle $\theta(t)$ of a magnetic wire versus time under a rotating magnetic field, illustrating the synchronous and asynchronous regimes according to the Maxwell model.* ***e)*** *Average angular velocity $\Omega(\omega/\omega_C)/\omega_C$ computed from Eq. 6. The shaded area indicates the frequency range in which the magnetic wire is synchronous with the field. The orange dots labelled c and d correspond to the $\theta(t)$ behaviors in Figs. 5c-d, respectively.* ***f)*** *Variation of wire oscillation angle $\theta_B(\omega)$ for the Maxwell model, together with the asymptotic value at high frequencies (adapted from Ref.[154]).* ***g)*** *Orientation angle $\theta(t)$ of a magnetic wire under a rotating magnetic field for the SLS model.* ***h,i)*** *Same as e) and f), respectively for the SLS model (adapted from Ref.[154]).*

**IV.3 – Recent applications of magnetic wire microrheology**

*IV.3.1 - Optical methods for two-dimensional viscosity mapping*

Two-dimensional viscosity mapping aims to quantify fluid viscosity with micrometric spatial resolution and to reconstruct real-time spatial distributions. In the context of biological fluids, approaches relying on molecular rotor probes, whose fluorescence wavelength depends on the viscosity of the surrounding medium[158] or on fluorescence polarization anisotropy[159,160] have been the most widely used in recent years. Their main limitation, however, is that the spatial resolution is constrained by optical diffraction, as these methods depend on ensembles of fluorophores. In contrast, solid nanoparticles present an appealing alternative: their strong optical scattering enables the super-localization of individual probes and thus the mapping of viscosity at scales below the diffraction limit.[161] For anisotropic particles such as needles or rods, the rotational dynamics can be exploited, allowing localized viscosity measurements to be performed over extended periods of time. Using sub-micron rods, orientation can be determined by optical microscopy and imaging. To achieve enhanced spatial resolution, sub-diffraction rod-like particles with highly anisotropic scattering efficiencies have recently been developed, such as Janus particles.[162] or gold nanorods.[163] Under directional illumination, these particles appear as single, non-resolved diffraction patterns. Owing to their geometry, the resulting figure is anisotropic and exhibits a characteristic toroidal distribution. The rotation of a single rod thus produces pronounced variations in the collected intensity, resulting in blinking patterns.[163,164] Monitoring the temporal variations in intensity thus provides a measurement of the rotation frequency and its dependence on the surrounding viscosity. In this section, we present two photonic techniques that use magnetic rods actuated by a magnetic field to measure viscosity, either at the diffraction limit or below. These methods are illustrated using model fluids and rely on the dialysis synthesis described in Section IV.1. The wire dispersions are then sonicated and magnetically filtered to obtain submicron structures, which we refer to as rods.

*Holographic detection of rotating nanorods*

A key advantage of magnetic rods is their ability to enable active measurements when driven by an external rotating field at fixed frequency. In this configuration, the field sets the rod rotation or oscillation frequency, depending on whether it operates in the synchronous or asynchronous regime, as described in Section IV.2. In general, cameras mounted on optical microscopes are too slow to properly capture the kilohertz-range blinking frequencies exhibited such probes. In this case, stroboscopy can be implemented by modulating both the illumination and the reference waves with acousto-optic modulators, thereby reducing the hologram modulation frequency. Combined with multiplexed lock-in detection and numerical demodulation,[165] this approach also significantly enhances the signal-to-noise ratio. The holographic interference system described in Fig. 6a allows the sensitive detection of individual rods, and holographic reconstruction allows 2D or 3D maps of the blinking intensity for frequency $\omega$ in the range 1-$10^6$ rad s$^{-1}$. Figs. 6b and 6c display typical blinking spectra *versus* frequency obtained on non-resolved rods submitted to a rotating field for water-glycerol mixtures with viscosities 12 and 2.6 mPa s respectively. The intensity drops sharply with increasing $\omega$, with the transition between the two regimes occurring at $\omega_C$. Plots of the critical frequency measured in water-glycerol mixtures of viscosities ranging from $10^{-3}$ to $10^{-2}$ Pa s display a linear relationship with $1/\eta_0$, in agreement with Eq. 5. These results demonstrate that $\omega_C$ can be determined with holographic detection, and also used to measure viscosity maps (Fig. 6d).

*Towards sub-diffraction viscosity imaging*
The same holographic detection system also offers the possibility to detect and localize individual nanoparticles in 3D with precisions reaching now a few tens of nanometers. When the particles are sufficiently diluted to be statistically separated by more than one optical point spread function (PSF), here typically 2 µm, the 3D image of each individual rod appears as a single PSF. Calculating the center of mass of these PSFs enables superlocalization with nm-scale precision. The spectra shown in Fig.6e are obtained using a single magnetic rod, and display a transition frequency $\omega_C$ which is correlated to the viscosity of the medium. If the total measurement time can be kept short, typically on the order of a few seconds, limiting particle diffusion, the explored volume can be as small as 0.5 µm³. This enables viscosity measurements with spatial resolutions of 10–100 nm well below the optical diffraction limit.[166]

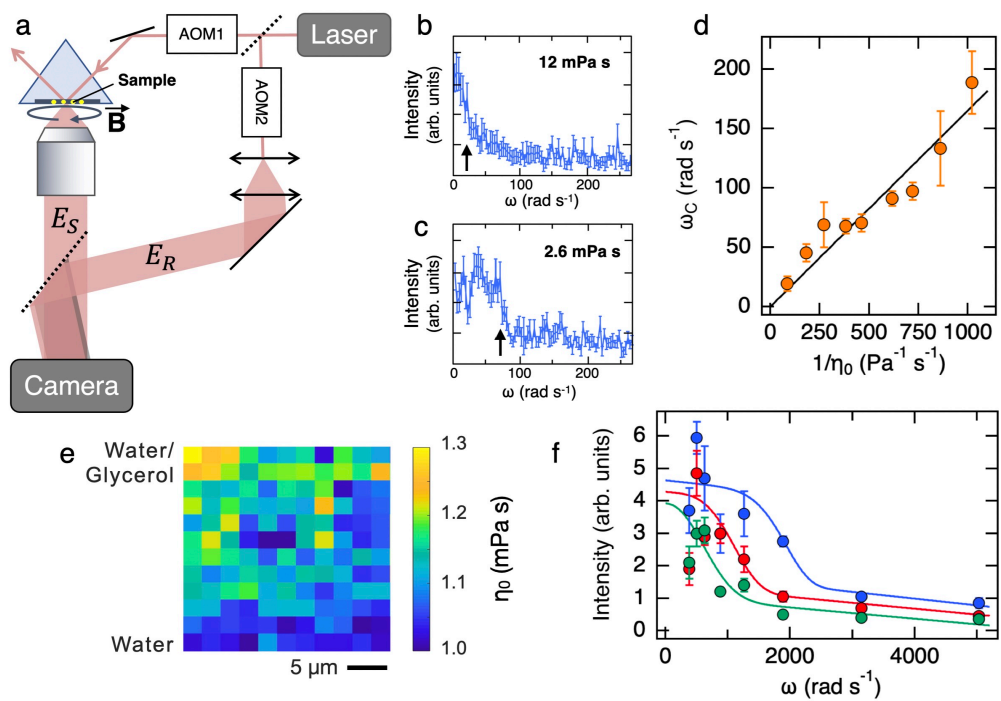

*Figure 6: a) Schematic representation of a digital holography set-up for the dark-field imaging of nanoparticles under total internal reflection illumination. Acousto-optical modulators (AOM) are powered in the MHz range, allowing the detection of fast blinking. Using the total internal reflection scheme, most of the illumination laser is rejected to avoid blinding the camera, and only the blinking light scattered by the*

*particles is collected.* ***b and c)*** *Blinking spectra obtained from magnetic rods with increasing frequency for water/glycerol with viscosities 12 and 2.6 mPa s respectively. The rods were synthesized from MNPs using the dialysis method and subsequently sonicated to reduce their length to 1.0 μm, with a diameter of 0.44 μm. The arrows indicate $\omega_C$.* ***d)*** *Critical frequency $\omega_C$ plotted against the inverse zero-shear viscosity in water–glycerol mixtures. The straight line corresponds to Eq. 5 and was obtained without adjustable parameters.* ***e)*** *Viscosity gradient measured by determining $\omega_C$ in each pixel.* ***f)*** *Blinking spectra obtained on single rods in liquids of various viscosities for a field modulation amplitude 3.5 mT. Fig. 6a-f are adapted from Ref.[166]*

*IV.3.2 – Intracellular microrheology of metastatic cancer cells*

About two decades ago, it was hypothesized that metastatic cancer cells - those capable of forming secondary tumors at distant sites, are softer than their healthy counterparts.[167,168] The rationale is that cells with high metastatic potential must cross multiple physical barriers, including the basal membrane surrounding the tumor, the ECM and the endothelium of a nearby vessel, before disseminating and colonizing new tissues. Increased deformability was therefore proposed as a key physical trait invading metastasis.[169,170] Based on this hypothesis, numerous groups have investigated cancer cell mechanics, most often using atomic force microscopy to probe whole-cell responses at the 10–30 μm length scale.[31] A broad range of cancer cell types has been also examined, including breast, pancreatic, bladder, prostate, and ovarian cancers.[32] Among them, a standard set of three human epithelial breast cell lines is widely used: the non-tumorigenic MCF-10A, the low-metastatic MCF-7, and the highly metastatic MDA-MB-231. Distinguishing between the latter two from a mechanical perspective is particularly important, as it is now established that nearly 90% of cancer-related deaths result from metastases rather than primary tumors.[171]

Fig. 7a compiles AFM measurements of the apparent Young modulus $E_{App}$ across these three lines, collected from 15 independent studies. Average values decrease from MCF-10A ($E_{App}$ = 3202 ± 2037 Pa) to MCF-7 ($E_{App}$ = 2039 ± 1339 Pa) and MDA-MB-231 ($E_{App}$ = 2105 ± 1297 Pa). However, the spread of values is wide, and the differences are not statistically significant. The variability found in AFM is often attributed to differences in experimental conditions, particularly indentation depth and location, cell morphology, or probe geometry.[172,173] Moreover, the apparent Young modulus alone does not allow a clear distinction between weakly and strongly metastatic cells. To probe mechanics at the cell scale, we used MRS with micron-sized magnetic wires internalized into the cytoplasm to measure the zero-shear viscosity and elastic modulus.[39] The strength of MRS lies in its ability to probe very low frequencies (down to $10^{-3}$ rad s$^{-1}$), enabling direct $\eta_0$-determination from the critical frequency separating synchronous and asynchronous regimes. As shown in Fig. 7b, intracellular viscosity varies significantly between cell lines, with median values $\eta_0$ = 36.3 ± 11.2 Pa s for MCF-10A, 65.9 ± 11.4 Pa s for MCF-7, and 12.0 ± 5.7 Pa s for MDA-MB-231. In the figure, each data point corresponds to a single wire in a different cell. Notably, statistical analysis confirmed that MDA-MB-231 cells have a viscosity 3 times lower than that of MCF-10A and nearly six times lower than that of MCF-7. These findings suggest that, beyond increased deformability, intracellular viscosity may constitute a key mechanical marker of metastatic potential. Lower viscosity accelerates cellular dynamics, enhancing the ability of tumor cells to cross multiple barriers during dissemination.[31] Extending MRS investigations to other cancer types should help validate these initial findings and advance both fundamental mechanobiology and biomedical applications.

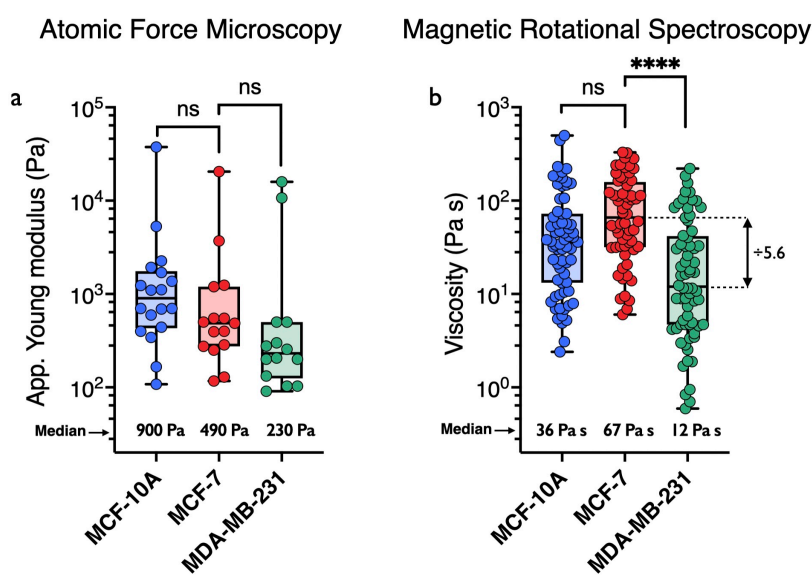

*Figure 7: a) Meta-analysis of atomic force microscopy (AFM) indentation measurements on normal (MCF-10A) and cancerous (MCF-7 and MDA-MB-231) breast epithelial cells. Data compiled from 15 independent studies provide statistical distributions of the apparent Young modulus ($E_{App}$) at the whole-cell scale. b) Zero-shear viscosity boxplots for MCF-10A, MCF-7 and MDA-MB-231. The values of the median viscosity are indicated (adapted from Ref.[39]).*

## IV – Conclusion

Since the discovery by René Massart of the coprecipitation method for producing iron oxide nanoparticles, extensive efforts have been devoted to their applications in life sciences. The magnetic moment of iron oxide nanocrystals, and the local magnetic field it generates, enable their use in a broad range of magnetism-based applications, such as MRI, magnetic separation, targeted drug delivery, or hyperthermia. While some of these technologies have reached maturity, others remain under active development or have yet to demonstrate their translation potential. This review examines the use of Massart nanoparticles in mechanobiology, a discipline that investigates how the mechanical properties of cells and tissues regulate biological processes. Mechanobiology has attracted increasing interest in recent years, particularly in the context of cell and tissue rheology. A prerequisite for the successful integration of nanomaterials into biological systems, and MNPs are no exception to this rule, is the precise control of interfacial properties. This control relies on advanced surface chemistry, including the use of functional polymers that maintain colloidal stability and minimize toxicity. Equally important is the ability to synthesize reproducible and scalable quantities of nanoparticles. The Massart coprecipitation method, combined with high-throughput coating strategies, satisfies these requirements and supports extensive screening of experimental conditions typical of biological investigations. The first part examines MNP assemblies that form spontaneously during internalization within the cytoplasm in a wide range of cell types, including immune, endothelial, epithelial, stem and cancer cells. These assemblies are first employed to promote the formation of cellular aggregates and subsequently to apply homogeneous magnetic forces for investigating tissue deformation and viscoelastic response. From a rheological point of view, measurements on cellular aggregates exhibit characteristics typical of viscoelastic materials and Young moduli ranging from 10 to 100 Pa, indicating soft and deformable biomaterials. We show, for example, that this approach makes it possible to perform creep experiments on millimeter-sized spheroids and to determine their elastic responses. The second part focuses on another type of MNP assembly, namely superparamagnetic wires, which are used as probes for complex fluid active microrheology. In this context, rotational magnetic spectroscopy is especially suitable for living cells, providing quantitative measurements of cytoplasmic viscosity, which typically ranges from 10 to 100 Pa s across multiple cell lines. From a rheological standpoint, it is shown that the intracellular medium behaves as a viscoelastic liquid characterized by a defined viscosity, elastic modulus, and relaxation time. Further developments have enabled the use of these wires to generate viscosity mapping through advanced photonic and holographic techniques. As this review highlights, Massart magnetic nanoparticles have long proven their versatility in diverse realms of science, and we show here that they continue to offer new perspectives for the mechanical study of cells and tissues

## Author contributions
Conceptualization, Methodology, Validation, Formal analysis, Investigation, Visualization, Writing – original draft, Funding Acquisition: MR, GT, JFB
Supervision, Writing – review and editing: JFB

## Conflict of interest
The authors have no competing interests to declare

## Ethics approval
Ethics approval is not required

## AI Disclosure Statement
Artificial intelligence was used to improve the clarity and English language of selected sentences and paragraphs. No AI-generated content was used for data analysis, scientific interpretation, or drafting original research findings. All scientific content was conceived, written, and validated by the authors.

# Data Accessibility Statement

The data presented in this review are available through the corresponding ZENODO repository link: https://doi.org/10.5281/zenodo.17707567

# Acknowledgments

We thank Grégory Arkowitz, Claude Bostoen, Anna Dubrovska, Clémence Gentner, Robert Kuszelewicz, Benoit Ladoux, René-Marc Mège and Milad Radiom for fruitful discussions. MR and JFB also express their gratitude to the PHENIX laboratory at Sorbonne University, particularly to Jérôme Fresnais and Christine Ménager, for providing the iron oxide nanoparticles synthesized by the Massart method throughout these years. This research was supported in part by the Agence Nationale de la Recherche under the contracts ANR-21-CE19-0058-1 (MucOnChip) and ANR-24-CE42-6142-03 (ViscoMag2).

# List of abbreviations

| | |
|---|---|
| 2D | Two-dimensional |
| 3D | Three-dimensional |
| AFM | Atomic Force Microscopy |
| ECM | Extracellular Matrix |
| EDX | X-ray energy-dispersive spectroscopy () |
| MNP | Massart Magnetic Nanoparticles |
| MRS | Magnetic Rotational Spectroscopy |
| PAA | Poly(acrylic acid) |
| PEG | Poly(ethylene glycol) |
| PSF | Point Spread Function |
| SEM | Scanning Electron Microscopy |
| SLS | Standard Linear Solid model |
| TEM | Transmission Electron Microscopy |
| VSM | Vibrating Sample Magnetometry |

# TOC Image

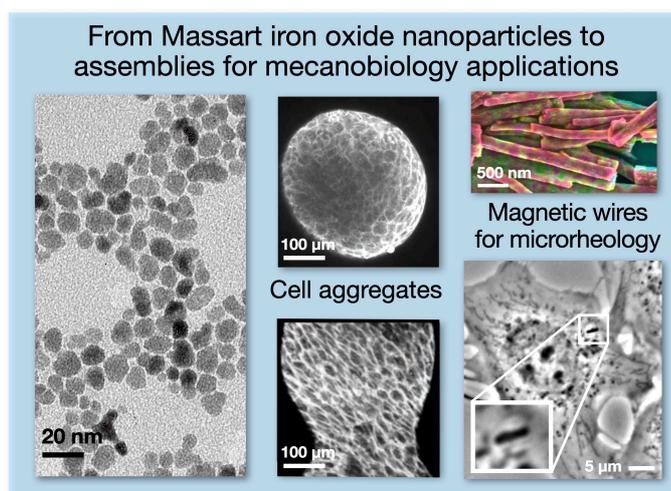